\documentclass[twocolumn]{revtex4}

\usepackage{graphicx}
\usepackage{dcolumn}
\usepackage{amsmath}

\begin{document}

\title{Phase diagram and mechanism of superconductivity
in a strongly correlated electron system}

\author{Takashi Yanagisawa$^1$, Mitake Miyazaki$^2,$ Kunihiko Yamaji$^1$
}

\affiliation{
$^1$ Electronics and Photonics Research Institute,
National Institute of Advanced Industrial Science and Technology,
1-1-1 Umezono, Tsukuba, Ibaraki 305-8568, Japan\\
$^2$Hakodate Institute of Technology, 14-1 Tokura, Hakodate,
Hokkaido 042-8501, Japan
}

\begin{abstract}
We investigate the phase diagram of two-dimensional (2D) Hubbard model 
by employing the optimization variational Monte Carlo method.
The 2D Hubbard model is the most simple electronic model for
cuprate high-temperature superconductors.
The phase diagram consists of three regions; they are antiferromagnetic
insulator (AFI) region, superconducting (SC) region and the coexistent region of
superconductivity and antiferromgantism.
The phase diagram obtained by numerical calculations well agrees with
the experimental phase diagram for high-temperature cuprates.
We mainly focused on the effect of $t'$ on the antiferromagnetic (AF) correlation
and the AFI region.  The area of the AF phase increases when we include $t'$ and 
thus the pure $d$-wave SC phase decreases.  The AFI phase near half filling 
decreases as $|t'|$ increases.
\end{abstract}

\maketitle

\section{Introduction}
\label{intro}

The mechanism of high-temperature superconductivity has been
studied intensively for more than 30 years\cite{bed86}.
It is important to clarify the phase diagram of strongly correlated
electron states in the study of high-temperature 
superconductivity\cite{agr03,fra10,kei15,cam15,ryb16}.
The purpose of this study is to understand the phase diagram of cuprate
high-temperature superconductors since the mystery of the phase 
diagram in cuprates has never been resolved.
The electron correlation between electrons plays an important role
in cuprate superconductors because the parent materials without
carrier doping are Mott insulators and the Cooper pairs have the
$d$-wave symmetry.
It is very important to clarify the electronic properties of electrons
in the CuO$_2$ plane\cite{eme87,hir89,sca91,web09,lau11,yan01,yan03,yan01a}.
The model for  the CuO$_2$ plane has $d$ electron in copper atoms
and $p$ electrons in oxygen atoms.
We often examine the simplified model, by neglecting oxygen
sites in the CuO$_2$ plane, which is called the (single-band) Hubbard
model\cite{hub63,hir85,yok88,mor85,yos96}.
It is an important subject whether the two-dimensional (2D) Hubbard model
has a superconducting phase or 
not\cite{zha97,zha97b,aim07,bul02,miy04,yan08,miy09,yan09,yam11,yan13}.
The variational wave functions have been improved intensively
recently\cite{yok06,yok13,yok18,yan16,yan19,yan19b}.
Recent results on the 2D Hubbard model are now supporting
the existence of superconductivity in the ground 
state\cite{yok18,yan16,yan19,yan19b}.

A variational Monte Carlo method is a suitable method to
investigate electronic properties of strongly correlated electron
systems\cite{yok88,nak97,yam98}.
A variational wave function is improved and optimized by introducing
new variational parameters to control the electron correlation.
We have proposed correlated wave functions by multiplying an
initial wave function by $\exp(-S)$-type operators\cite{yan16,yan19,ots92,yan98},
where $S$ is a correlation operator.  The wave function is
further optimized in a systematic way by multiplying by the
exponential-type operators repeatedly\cite{yan16}. 
The ground-state energy evaluated by our wave function is much lower than
that by previous wave functions.

\section{Model Hamiltonian}

The CuO$_2$ plane consists of oxygen atoms and copper atoms (shown in Fig. 1).
The basic model for this plane is the three-band d-p model which
explicitly contains both oxygen $p$ electrons and copper $d$ electrons.
When we neglect oxygen atoms in this model, we have
the two-dimensional Hubbard model that consists of only $d$ electrons
on a lattice shown in Fig. 2.
The Hubbard model is given as
\begin{equation}
H = \sum_{ij\sigma}t_{ij}c_{i\sigma}^{\dag}c_{j\sigma}
+U\sum_i n_{i\uparrow}n_{i\downarrow},
\end{equation}
where $\{t_{ij}\}$ are transfer integrals and $U(>0)$ is the on-site
Coulomb energy.  The transfer integral $t_{ij}$
for nearest-neighbor pairs $\langle ij\rangle$ is denoted as
$t_{ij}=-t$ and that for next-nearest neighbor pair
$\langle\langle ij \rangle\rangle$ is $t_{ij}=-t'$.
Otherwise, $t_{ij}$ vanishes.
We denote the number of sites as $N$ and the number of electrons
as $N_e$.  The energy unit is given by $t$.
$n_{i\sigma}$ refers to the number operator:
$n_{i\sigma}= c^{\dag}_{i\sigma}c_{i\sigma}$.
The second term in the Hamiltonian represents the on-site
repulsive interaction between electrons with opposite spins.

One may understand the appearance of inhomogeneous states reported
for high-temperature cuprates based on the Hubbard 
model\cite{miy09,tra96,bia96,bia13}.
Concerning the existence of superconducting phase in the 2D
Hubbard model, quantum Monte Carlo studies have given negative
results and do not support high-temperature superconductivity
in the Hubbard model\cite{zha97,zha97b,aim07}.
The recent results based on elaborated optimized wave functions,
however, have provided a support for
superconductivity\cite{yan16,yan19}, especially in the strongly
correlated region\cite{yan16}.
In our opinion, it seems evident that the 2D Hubbard model
has a superconducting phase in the ground state.
There is, however, still an issue that should be clarified.
This is the competition between superconducting and
antiferromagnetic states.

\begin{figure}
\begin{center}
  \includegraphics[width=5.6cm]{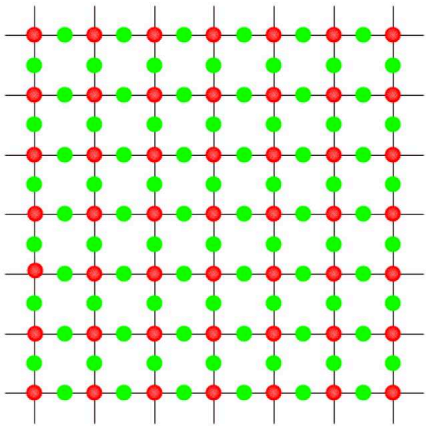}
\caption{Lattice of the CuO$_2$ plane with oxygen and copper atoms.
}
\label{cuo2}       
\end{center}
\end{figure}

\begin{figure}
\begin{center}
  \includegraphics[width=5.6cm]{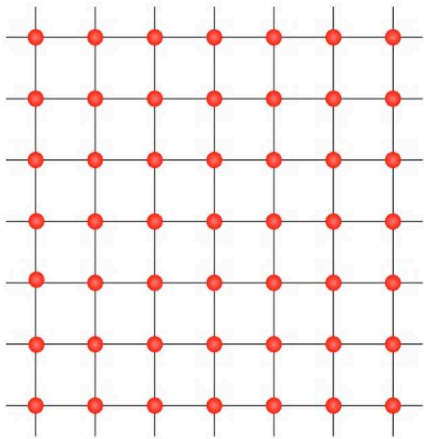}
\caption{Lattice of copper atoms in the CuO$_2$ plane.
}
\label{dlattice}       
\end{center}
\end{figure}

\section{Optimized Wave Functions}

We use an ansatz for the wave function and evaluate the expectation
values using the variational Monte Carlo method.
A starting wave function to take account of the electron correlation
is given by the Gutzwiller ansatz:
\begin{equation}
\psi_G = P_G\psi_0 ,
\end{equation}
where $P_G$ is the Gutzwiller operator
$P_G=\prod_j(1-(1-g)n_{j\uparrow}n_{j\downarrow})$ with the
parameter $g$ in the range of $0\le g\le 1$.
$\psi_0$ is a trial function of one-particle state.
To investigate a stability of the superconducting (SC) state, we use
the BCS wave
function $\psi_{BCS}$ for $\psi_0$ with the gap parameter $\Delta_{SC}$.
The condensation energy is defined as
$E_{cond}=E(\Delta_{SC}=0)-E(\Delta_{SC}=\Delta_{SC,opt})$ for the optimized
gap function $\Delta_{SC,opt}$.
The antiferromagnetic (AF) one-particle state $\psi_{AF}$ is given
by the eigenstate of the AF trial Hamiltonian given by
\begin{equation}
H_{AF}= \sum_{ij\sigma}t_{ij}c^{\dag}_{i\sigma}c_{j\sigma}
-\Delta_{AF}\sum_{i\sigma}\sigma (-1)^{x_i+y_i}n_{i\sigma},
\end{equation}
where ${\bf r}_i=(x_i,y_i)$ are the coordinates of the site $i$.
$\Delta_{AF}$ indicates the AF order parameter.

Our wave function is obtained by multiplying $\psi_G$ by an off-diagonal
correlation operator to take account of intersite correlation.
The wave function is written
as\cite{yan16,ots92,yan98,yan99,eic07,bae09,bae11,bae19}
\begin{equation}
\psi_{\lambda} = e^{-\lambda K}\psi_G,
\end{equation}
where $K$ indicates the kinetic term of the Hamiltonian
$K=\sum_{ij\sigma}t_{ij}c^{\dag}_{i\sigma}c_{j\sigma}$ and
$\lambda$ is a real constant which is the variational
parameter chosen to lower the ground-state energy\cite{yan16,yan19,yan98,yan99}.
The initial wave function $\psi_0$ is written by Slater determinants
in the real space representation.  The basis states are given by
Slater determinants.  
The operator $e^{-\lambda K}$ 
produces off-diagonal elements between different basis states
and lowers the ground-state energy.

This wave function is easily generalized to multi-band models and
appears to be a good wave function for the three-band d-p 
model\cite{yan19,yan14}.

\section{Phase diagram and the effect of $t'$}

We consider the phase diagram as a function of the doping rate
(hole density) and examine the effect of the nearest-neighbor transfer
$t'$ on it.
The common feature of the phase diagram of high-temperature cuprates
is that the antiferromagnetic insulator phase exists when the hole
density $x$ is small near half-filling and there is the $d$-wave
SC phase when the hole density is larger than the
critical value $x_c$ where $x_c\sim 0.05$.  The SC phase vanishes when 
$x$ becomes as large as about 0.25.
 
We show the phase diagram in Fig. 3 where the condensation energy
is shown as a function of the hole density $x$.  The condensation
energy is defined as the energy difference given as
\begin{equation}
\Delta E = E(\Delta=0)-E(\Delta=\Delta_{opt}),
\end{equation}
where $\Delta$ is the order parameter for SC ($\Delta_{SC}$) and 
AF ($\Delta_{AF}$) and $\Delta_{opt}$
indicates the optimized value of $\Delta$.
In Fig. 3 calculations were performed for $U/t=18$ and $t'=0$ on
a $10\times 10$ lattice and we include the results for the AF state for
$U/t=14$ and 12.
There occurs the phase separation when $x<0.06$, and the AF state
in this region is an insulating state\cite{yan19,yan19b}.
When $x>0.06$, the ground state becomes $d$-wave superconducting.
In the region near the AF boundary given as $0.06<x<x_{dSC}$,
the AF order and superconductivity coexist where $x_{dSC}$ is
approximately $x_{dSC}\sim 0.08-0.09$.
The pure $d$-wave state is realized for $x>x_{dSC}$.
When $U$ decreases, the AF order parameter increases.

We investigate the effect of $t'$ here.
The AF condensation energy is shown as a function of the hole density 
$x$ in Fig. 4 for $t'/t=-0.1$ and $U/t=18$.  The AF region becomes
large as $-t'$ increases, and thus
the $t'=0$ is most favorable for the pure $d$-wave SC state.
The phase-separation region decreases, that is,
the antiferromagnetic insulator (AFI) region decreases due to $t'$.
The Fig. 5 presents the phase-separation (PS) region on the $x$-$t'$ plane.
The PS region disappears when $t'$ is as large as $-0.2$.

To summarize the results, the $t'$ increases antiferromagnetic correlation
and decreases insulator region near half-filling.

\begin{figure}
\begin{center}
  \includegraphics[width=7.2cm]{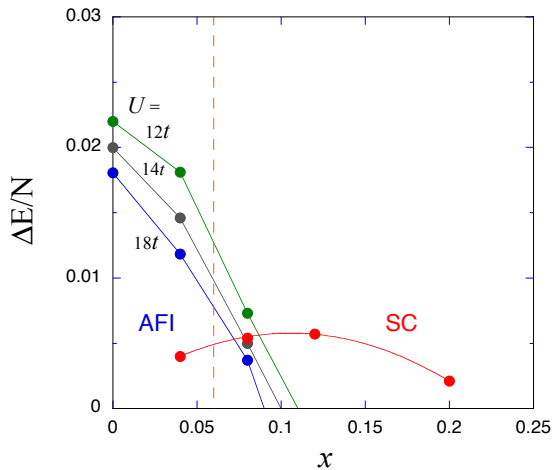}
\caption{
The condensation energy as a function of the hole density $x$
for $t'=0$.
}
\label{de1}       
\end{center}
\end{figure}

\begin{figure}
\begin{center}
  \includegraphics[width=7.2cm]{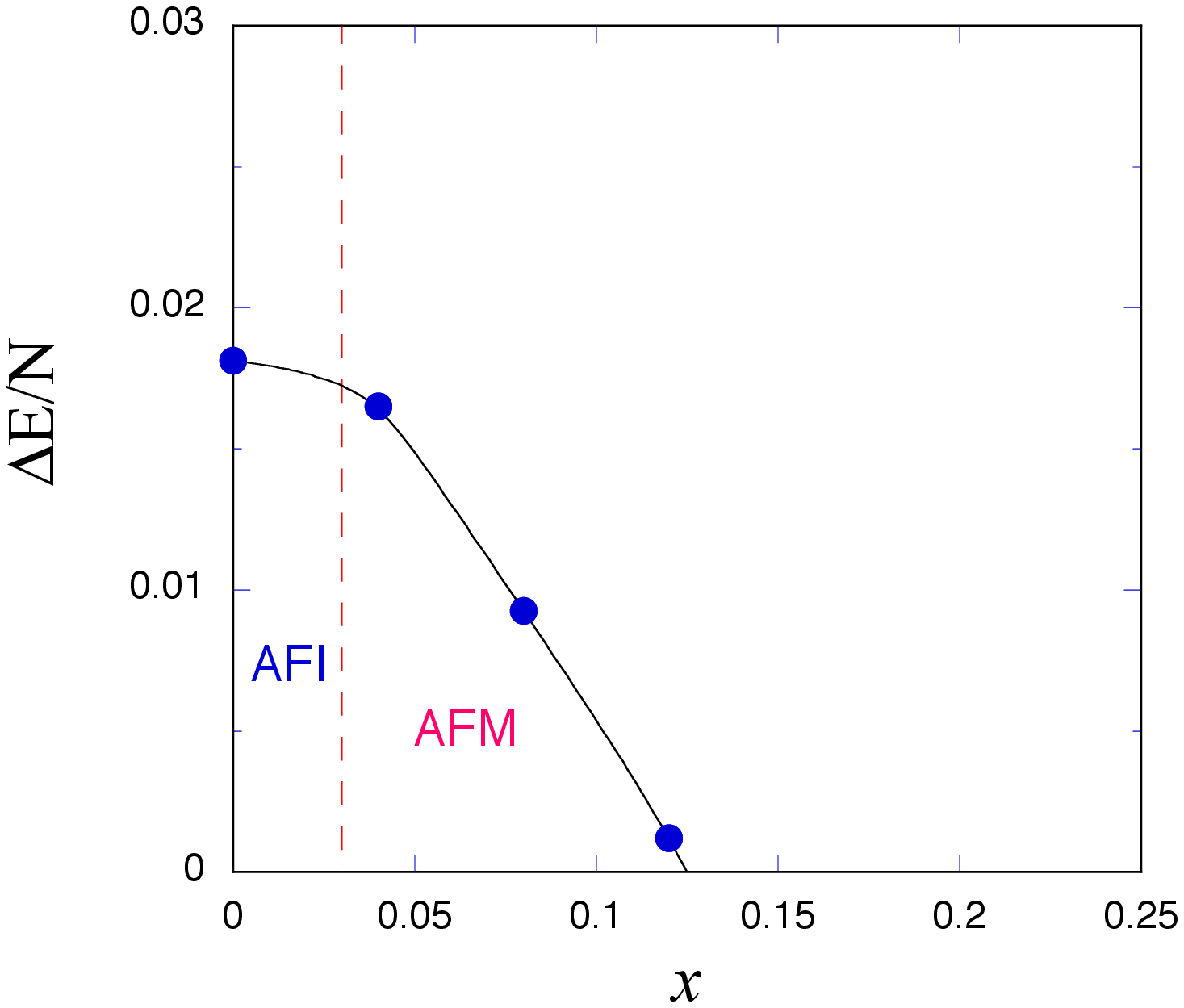}
\caption{
The condensation energy as a function of the hole density $x$
for $t'=-0.1$.
}
\label{de2}       
\end{center}
\end{figure}

\begin{figure}
\begin{center}
  \includegraphics[width=7cm]{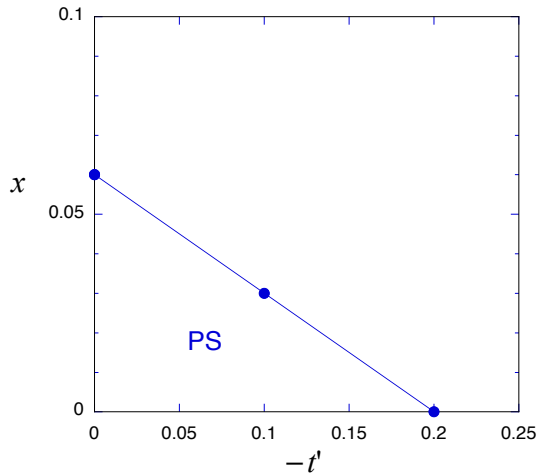}
\caption{
Phase-septration region in the phase-septration region in the
plane of $x$ and $t'$.
}
\label{xt2}       
\end{center}
\end{figure}

\section{Summary}

We have investigated the ground state of the 2D Hubbard model on the
basis of the optimization variational Monte Carlo method.
We employ the optimized wave function by introducing new
variational parameters to go beyond the Gutzwiller ansatz.
The ground-state energy is greatly lowered by our wave function.
The metal-insulator transition is also well described by our
optimized function\cite{yan14}.

We mainly focused on the effect of $t'$ on the AF correlation
and the phase separation. 
The area of the AF phase increases when we include $t'$ and thus the pure
$d$-wave SC phase decreases.  The AFI phase near half filling
decreases as $|t'|$ increases.

The electron pairing interaction is induced due to spin and charge
fluctuations in the strongly correlated region.
There is a crossover between weakly and strongly correlated
region as the strength of the Coulomb interaction $U$ increases.
Crossover phenomena have also been investigated in the study
of cuprate superconductors\cite{kag19}.
This kind of crossover may be universal which occurs with a
singularity in the intermediate region as in the Kondo effect,
QCD and BCS-BEC crossover\cite{kon12,yan12,ell96,noz85}.
The kinetic energy induced by the operator $\exp(-\lambda K)$
may drive the electron pairing and helps to bring about
high-temperature superconductivity.

\begin{acknowledgements}
The authors express their sincere thanks to Prof. D. Baeriswyl
for valuable discussions.
The computations were supported by the Supercomputer Center
of the Institute for Solid State Physics, the University of
Tokyo.
This work was supported by a Grand-in-Aid for Scientific
Research from the Ministry of Education, Culture, Sports,
Science and Technology of Japan (Grant No. 17K05559).
\end{acknowledgements}

\end{document}